\documentclass[twocolumn,amsmath,amssymb]{revtex4} 

\usepackage{graphicx}
\usepackage{dcolumn}
\usepackage{bm}

\begin{document}

\title{Lithographic engineering of anisotropies in (Ga,Mn)As}

\author{S. H\"{u}mpfner}
\affiliation{Physikalisches Institut (EP3), Universit\"{a}t W\"{u}rzburg,
Am Hubland, D-97074 W\"{u}rzburg, Germany}

\author{K. Pappert}%
\affiliation{Physikalisches Institut (EP3), Universit\"{a}t W\"{u}rzburg,
Am Hubland, D-97074 W\"{u}rzburg, Germany}

\author{J. Wenisch}%
\affiliation{Physikalisches Institut (EP3), Universit\"{a}t W\"{u}rzburg,
Am Hubland, D-97074 W\"{u}rzburg, Germany}

\author{K. Brunner}%
\affiliation{Physikalisches Institut (EP3), Universit\"{a}t W\"{u}rzburg,
Am Hubland, D-97074 W\"{u}rzburg, Germany}

\author{C. Gould}%
\affiliation{Physikalisches Institut (EP3), Universit\"{a}t W\"{u}rzburg,
Am Hubland, D-97074 W\"{u}rzburg, Germany}

\author{G. Schmidt}%
\affiliation{Physikalisches Institut (EP3), Universit\"{a}t W\"{u}rzburg,
Am Hubland, D-97074 W\"{u}rzburg, Germany}

\author{L.W. Molenkamp}%
\affiliation{Physikalisches Institut (EP3), Universit\"{a}t W\"{u}rzburg,
Am Hubland, D-97074 W\"{u}rzburg, Germany}

\author{M. Sawicki}
\affiliation{Institute of Physics, Polish Academy of Sciences,
al.~Lotnik\'ow 32/46, PL-02668 Warszawa, Poland}

\author{T. Dietl}%
\affiliation{Institute of Physics, Polish Academy of Sciences,
al.~Lotnik\'ow 32/46, PL-02668 Warszawa, Poland}

\date{\today}

\begin{abstract}
The focus of studies on ferromagnetic semiconductors is moving from
material issues to device functionalities based on novel phenomena
often associated with the anisotropy properties of these materials.
This is driving a need for a method to locally control the
anisotropy in order to allow the elaboration of devices. Here we
present a method which provides patterning induced anisotropy which
not only can be applied locally, but also dominates over the
intrinsic material anisotropy at all temperatures.
\end{abstract}

\maketitle

The coupling of transport and magnetic properties in ferromagnetic
semiconductors gives rises to many interesting anisotropy related
transport phenomena such as strong anisotropic magnetoresistance
(AMR), in-plane hall \cite{Roukes}, tunneling anisotropic
magnetoresistance (TAMR) \cite{GouldTAMR,ChrisTAMR} and Coulomb
blockade AMR \cite{CBAMR}. Studies on all of these effects so far
have primarily made use of the intrinsic anisotropy present in the
host (Ga,Mn)As layer. Before they can be harnessed to their full
potential, a means of engineering the anisotropy locally is needed,
such that multiple elements with different anisotropies can be
integrated, and their interactions can be properly investigated.

One successful approach to local anisotropy control in metallic
ferromagnets has been to make use of shape anisotropy. The same
approach has been tried in the prototypical ferromagnetic
semiconductor (Ga,Mn)As with lackluster results. In
Ref.~\onlinecite{Hamaya_06}, the authors reported the observation of
shape induced anisotropy in (Ga,Mn)As wires of 100 nm thickness x
1.5 x 200 $\mu m^2$, but only over a limited temperature range.
Moreover, our own experience in attempting to use wires of similar
dimensions have yielded sporadic results with the wires having
irreproducible anisotropy, with either biaxial or uniaxial easy axes
in inconsistent directions.

Furthermore, a simple calculation of the expected shape anisotropy
term in such wires indicates that it should not play a significant
role. While the infinite rod model used in \cite{Hamaya_06} does
predict an appreciable shape anisotropy field given by
$\mu_0M_S/2$, where $M_S$ is the sample magnetization, it is not
applicable to structures which are much thinner than their lateral
dimensions. A more exact rectangular prism calculation
\cite{Aaroni_98} gives a 5 times weaker shape anisotropy with an
anisotropy energy density of 80 J/m$^3$ which is much too small to
compete with the typical crystalline anisotropy of ~3000 J/m$^3$
\cite{Wang_05,Hamaya_06} in this material.

\begin{figure}[b] \includegraphics{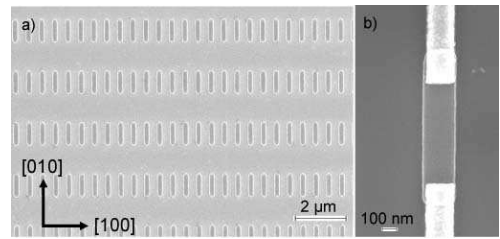}
\caption{\label{f1} \textbf{a}) SEM photograph of a small part of
a typical 8 million nanobar array. The individual bars have
lateral dimensions of 200 nm by 1 $\mu$m. \textbf{b}) An
individual nanobar contacted for transport characterization. }
\end{figure}

Growth strain reduces the cubic symmetry of the (Ga,Mn)As
zinc-blend crystal structure creating a uniaxial anisotropy with
an easy/hard magnetic axis in growth direction when the (Ga,Mn)As
layer is tensile/compressively strained. This growth strain is
known\cite{Furdyna} to influence the strength of
the perpendicular component of the anisotropy of the \emph{whole}
layer. Here we discuss (Ga,Mn)As grown on a (001) oriented GaAs
substrate, whose out-of-plane hard magnetic axis confines the
magnetization in the plane. Phenomenologically, the net in-plane
magnetic anisotropy is known to result from a competition of two
primary contributions: the crystal symmetry induced biaxial
anisotropy with [100] and [010] easy axes, and a uniaxial
anisotropy (the origin of which is not clear) with the direction
of its easy axis assuming either of the in-plane $\langle
110\rangle$ directions. The anisotropy of (Ga,Mn)As is further
enriched by the existence of a second uniaxial, which is very weak
and along [010] \cite{GouldTAMR}. The interplay of these 3
anisotropies, all of which depend on hole concentration \emph{p},
temperature \emph{T} \cite{Sawicki_05} and sample strain, leads to
a material with a very sophisticated anisotropy which can be
difficult to control or reproduce from one layer to the next.

In this letter we suggest that an additional agent, i.e.,
lithographically induced strain relaxation, also plays a significant role in
nano-patterned structures and is the only reasonable means by
which to properly exercise \emph{local} control of the anisotropy in
(Ga,Mn)As. We demonstrate that patterning imposed relaxation
effects can not only be observed in ferromagnetic (Ga,Mn)As but
also that these effects can dominate the magnetic anisotropy in
the entire temperature range, up to the Curie temperature
$T_{c}$. Our findings pave the way to production of
samples with \emph{locally} designed anisotropy behavior.

A pair of nominally identical, high quality, 20 nm thick
Ga$_{0.96}$Mn$_{0.04}$As layers grown on a GaAs substrate
\cite{growth} with a $T_{c}$ of 70 K are chosen for these studies.
They are patterned into arrays of either [100] or [010] oriented
nanobars for magnetic investigation and equivalent individual
nanobars contacted for transport investigations. Fig.~\ref{f1}
shows SEM photographs of the above mentioned structures. Each
individual bar has lateral dimensions of 200~nm by 1~$\mu$m. The
full array of them is defined using electron beam lithography with
a negative resist. After developing, the defined pattern is
transferred into the (Ga,Mn)As layer using chemically assisted ion
beam etching (CAIBE). As many as 8 million nanobars are laid out
to provide sufficient total magnetization for the magnetic
anisotropy studies carried out in the variable temperature
superconducting quantum interference device (SQUID) magnetometer.

We investigate the magnetization \emph{m} vs.~\emph{H} dependencies
of the sample in applied magnetic fields of up to $\pm$100~mT for
the four major in-plane orientations. All spurious background
signals originating from the substrate and sample holders are
subtracted from the data presented.

\begin{figure}[t]
    \centering
        \includegraphics[width=8.5cm]{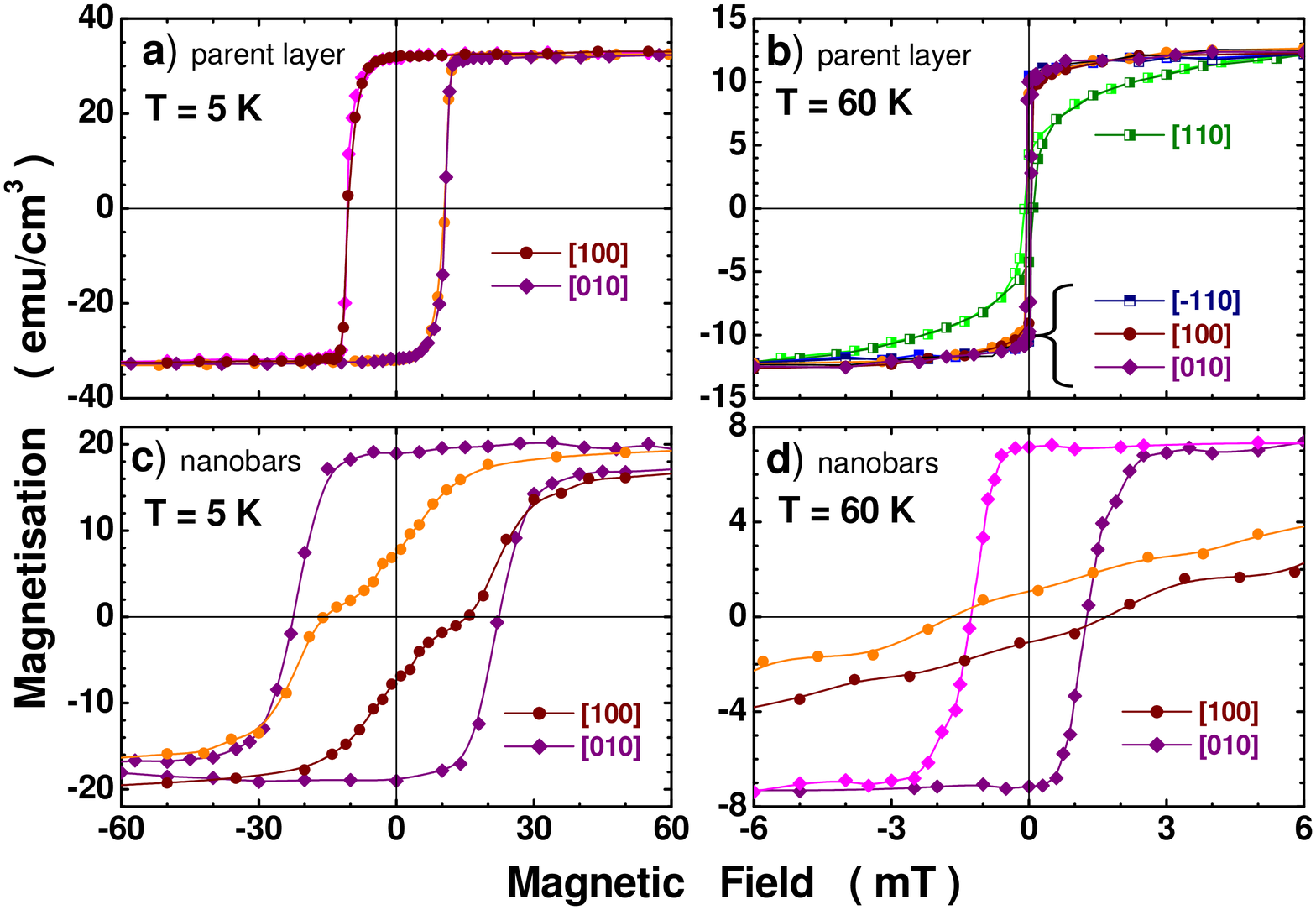}
    \caption{(color online) SQUID magnetization data for (\textbf{a-b}) the parent layer and
(\textbf{c-d}) the array of nanobars having their long side
aligned to [010].
  Light shades are used to mark the high resolution data obtained by numerical reflection to
  mimic the full hysteresis after confirming hysteretic symmetry with coarse measurements.
     }
    \label{F2}
\end{figure}

The salient features of the SQUID investigations are summarized in
Fig.~\ref{F2}, more elaborate magnetization studies of the
magnetic anisotropy in such devices will be published elsewhere
\cite{Sawicki_Ani-to-come}. We start with the unpatterned,
"parent" layer (top panels) to show that, as is typical for
(Ga,Mn)As, it exhibits equivalent behavior along [100] and [010],
both at very low \emph{T} (Fig.~\ref{F2}(\textbf{a})), and near
$T_{\mbox{\tiny C}}$ (Fig.~\ref{F2}(\textbf{b})). This is simply a
manifestation of the fact that the presence of a $\langle
110\rangle$ uniaxial anisotropy, which bisects the four-fold
$\langle 100\rangle$ easy directions and acts equivalently on
[100] and [010] does not break the symmetry between these
directions. The [010] uniaxial anisotropy is too weak to
measurably break the symmetry.

This behavior is in stark contrast to that of the patterned array,
as shown in the bottom panels of Fig.~\ref{F2}, where
magnetization studies of an array of nanobars oriented such that
their long axis is along the [010] direction are presented. This
axis is still a magnetic easy axis, similar to that of the host.
The magnetic response along the [100] direction, which is along
the short side of the nanobars has however been completely
modified, and now exhibits pronounced hard axis behavior. From the
hard axis measurements, we estimate the lithographically imposed
anisotropy field $\mu_0H_L$ produced by our sub-micron patterning
to be 25 and 20 mT at 5 and 60 K, respectively. This field is
comparable to the crystalline four-fold anisotropy field($\lesssim
100mT$) which dominates the behavior of the parent layer at 5 K,
and is much larger than the $\langle 110\rangle$ uniaxial term
($\sim 2 mT$) which dominates the behaviour of the unpatterned
layer at 60 K. For comparison, the shape anisotropy field is only
4 and 1.4 mT at these respective temperatures.

The overall magnetic anisotropy has thus been transformed from a
strongly temperature dependent mixture of four-fold and uniaxial
contributions into a well defined temperature independent uniaxial
behavior imposed locally along the long axis of the nanobars. Arrays
of nanobars patterned with their long axis along [100] give fully
equivalent results.

The submicron dimensions of the nanobars have also allowed us to
reach the single domain limit in (Ga,Mn)As at low temperatures. As
seen in Fig. \ref{F2}c, the magnetization reversal along the easy
axis of the nanobars takes place roughly at the uniaxial
anisotropy field of 25 mT, indicating a nearly fully coherent
behavior of the magnetization inside the nanobar. The situation is
more complicated at 60 K where, despite a 20-fold increase in the
coercivity of the nanobars compared to the parent layer, the easy
axis switch occurs at $\sim 1.5$ mT, which is only a small
fraction of the lithographically imposed anisotropy field of 20 mT
at that temperature. The reason for this is that the equivalence
of the \{[100], [010]\} and $[\overline{1}10]$ anisotropy energy
densities (Fig.~\ref{F2}b) observed in the parent layer at this
temperature facilitates magnetization rotation along the nanobar,
thus reducing their coercive field.

\begin{figure}[t]
\centerline{\includegraphics{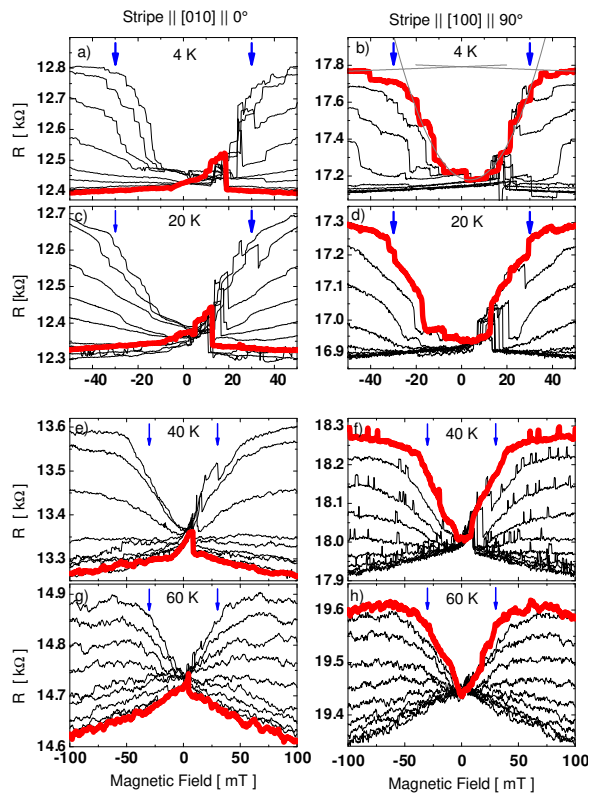}} \caption{\label{f3}
Magnetoresistance scans for angles between $0^\circ$ and
$90^\circ$ in 10$^\circ$ steps on the bars along [010] (left
column) and [100] (right) at various temperatures and 5 mV bias
voltage. The thick (red) line indicates the field sweep along the
[010] crystal direction. The arrows indicate the estimated
uniaxial anisotropy field.}
\end{figure}

Having achieved the desired anisotropy control in the arrays, we
now turn to electrical investigations, for which individual
nanobars are prepared using similar lithography as in the
patterning of the arrays. The major challenge in this case is a
non-perturbative way of contacting the nanobar. This is
non-trivial as it requires the formation of ohmic contacts onto
(Ga,Mn)As with a $\sim$100 nm length scale. Moreover, our
experience has shown that improperly optimized contacts do exert
strain onto the layer, significantly altering it's anisotropy
\cite{OurAdvMat}. We succeeded by using a Ti layer patterned by
lift-off as a mask. After etching, the Ti mask is removed, and
Ti/Au contacts are applied by e-beam lithography and lift-off.
This yields contacts with a resistance-area product of below
$10^{-6} \Omega \cdot cm^{2}$. In Fig.~\ref{f3} we present
transport characterization of two such nanobars patterned along
the [100] and [010] directions on the same chip.

This sample is cooled in a variable temperature cryostat fitted
with a vector field magnet, and its magnetoresistance (MR)
behavior is measured for magnetic field applied along various
angles $\varphi$ ($0^\circ$ along [010]) in the plane of the
layer. Prior to every scan, the sample is magnetized at -300 mT
along $\varphi$.

The observed behavior (Fig.~\ref{f3}) is due to AMR, i.e. the
fact, that the resistivity of a ferromagnetic material depends on
the angle between the current and its magnetization. In (Ga,Mn)As
the resistance is higher if the magnetization is perpendicular to
the current($R_{\bot}$), than if both are parallel ($R_{\|}$). One
can thus infer the angle $\vartheta$ between magnetization and
current from the resistance R at any field value through\cite{Jan}

\begin{equation}
R(\vartheta)=R_{\bot}-(R_{\bot}-R_{\|}) \cos^{2}\vartheta
\end{equation}

and from the magnetization behavior deduce the magnetic anisotropy
of the (Ga,Mn)As stripes. The left part of Fig.~\ref{f3} presents MR
scans on the nanobar along the [010] crystal direction at various
temperatures. A common feature is that field sweeps along the
nanobar axis ($0^\circ$, thick (red) line) yield a low resistance
curve in the plot, indicating, through Eq. 1 that $M$ remains
parallel to the nanobar throughout the field sweep. The lowest
resistance is observed also at zero external field, confirming that
the nanobar axis is the magnetic easy axis in the whole temperature
range.  When the field is swept through positive values, the
magnetization reversal process is seen at the same magnetic field
values as in the magnetization measurements.

When the field is swept perpendicular to the nanobar (highest
curves), the large values of the resistivity at high magnetic
fields confirm that the magnetization is forced perpendicular to
the nanobar. The resistance decreases monotonically as the field
is swept down to zero where all curves meet, indicating that the
magnetization always rotates to the same angle, i.e. to the
uniaxial easy axis direction, previously established to be along
the long axis of the nanobar. The linear slope seen to develop in
the high magnetic field range is the isotropic magnetoresistance
\cite{Matsukura_2004}.

The right part of Fig.~\ref{f3} presents results for the nanobar
oriented along [100]. Since the coordinate system is fixed to the
crystallographic axes, and not the axis of the nanobar, the fully
opposite MR properties clearly indicate that the uniaxial behavior
is related to the elongated shape of the nanobar. The parent layer
easy axis perpendicular to the wire has been overwritten by the
patterning process and the lithographically imposed uniaxial
anisotropy is the dominant anisotropy up to $T_{c}$, as was seen
in the magnetization investigations. Employing Eq. 1 allows us to
assess the strength of this anisotropy. The hard axis MR-scan
would be parabolic if only a pure uniaxial anisotropy was present.
In such a case the magnetic field necessary to force the
magnetization perpendicular to the easy axis is a direct measure
for the strength of the anisotropy. To estimate this anisotropy
field, we fit a parabola \cite{West_60} to the low field data of
the perpendicular field scan in Fig.~\ref{f3}b and interpolate the
isotropic magnetoresistance of this scan back to the origin (thin
grey lines). The fitted parabola is slightly shifted towards
positive fields, which indicates the presence of a small biaxial
anisotropy contribution. The intersections between the grey lines
and the parabola give $\mu_0H_{L}\sim30$ mT. The same number
(marked with blue arrows) is a reasonable estimate for
$\mu_0H_{L}$ in all parts of Fig.~\ref{f3} indicating that indeed
the lithography induced uniaxial anisotropy is almost unchanged
between 4 K and 60 K. This is a strong indication that the present
effect is fundamentally different from classic shape anisotropy,
which depends on the volume magnetization, and thus decreases with
increasing temperature until it vanishes at $T_{c}$. Moreover,
while size effects may play a role in the observed increase of the
coercive field, they would play no role in modifying the
anisotropy. Obviously, the results presented here do not provide
direct evidence of strain relaxation. Direct confirmation would
require x-ray diffraction measurement which are not possible on
the small structures investigated here. However, we have been able
to verify that strain relaxation is the important agent in the
effects reported here using x-ray diffraction measurements on long
and narrow etched (Ga,Mn)As stripes\cite{Wenisch}.

In conclusion, we have demonstrated a reliable technique for
comprehensively controlling the anisotropy locally in (Ga,Mn)As
using a lithographic technique. We believe this will prove itself a
useful tool for studying novel spintronics effects related to
transport between regions of different anisotropies or unique
magnetization configurations within a layer.

\begin{acknowledgments}
The authors wish to thank V. Hock and T. Borzenko for help in sample
fabrication, and to acknowledge financial support from the EU
(NANOSPIN: FP6-IST-015728) and the German DFG (BR1960/2-2). Two of
us (MS and TD) acknowledge JST (ERATO).
\end{acknowledgments}

\pagebreak

\end{document}